\documentclass{pasa}%
\usepackage{placeins}
\usepackage{soul}

\title[Student Research Project Review]{A Review of High School Level Astronomy Student Research Projects over the last two decades.}
\author[Fitzgerald et al.]{M.T. Fitzgerald $^1$, R. Hollow$^2$, L. M. Rebull$^3$, L. Danaia$^4$  \and D.H. McKinnon$^5$ \\
\affil{$^1$Department of Physics and Astronomy, Macquarie University, NSW, 2109, Australia}%
\affil{$^2$CSIRO Astronomy and Space Science, PO Box 76, Epping, NSW 1710, Australia}%
\affil{$^3$Spitzer Science Center/Caltech, M/S 314-6, 1200 East California Boulevard, Pasadena, CA 91125, USA}%
\affil{$^4$School of Teacher Education, Charles Sturt University, Bathurst, NSW 2795, Australia }%
\affil{$^5$School of Education, Edith Cowan University, Mount Lawley, WA 6050, Australia }}%
\jid{PASA}
\doi{10.1017/pas.\the\year.xxx}
\jyear{\the\year}

\begin{document}%
\begin{abstract}
Since the early 1990s with the arrival of a variety of new technologies, the capacity for authentic astronomical research at the high school level has skyrocketed. This potential, however, has not realized the bright-eyed hopes and dreams of the early pioneers who expected to revolutionise science education through the use of telescopes and other astronomical instrumentation in the classroom. In this paper, a general history and analysis of these attempts is presented. We define what we classify as an Astronomy Research in the Classroom (ARiC) project and note the major dimensions on which these projects differ before describing the 22 major student research projects active since the early 1990s. This is followed by a discussion of the major issues identified that affected the success of these projects and provide suggestions for similar attempts in the future. 
\end{abstract}
\maketitle%
\section{INTRODUCTION }
\label{sec:intro}

Student research, pitched at an appropriate level,  can be an effective approach to address the disenchantment about science that results from traditional 'chalk'n'talk' styles of teaching (Hollow 2000). While students can be excited by the breathtaking images in astronomy alone, through research a student can gain a sense of pride and ownership in their work, as well as gain useful secondary meta-skills such as organisational techniques and problem-solving approaches that become invaluable in later studies and work. In addition, some students demonstrate greater capabilities when they are exposed to study at depth than one would have previously expected from their in-class behaviour. Gifted students as well as underachievers and the easily bored can be motivated by tackling some of the 'big questions' that a typical everyday science class lacks (Hollow 2005). 

It is also the case that traditional styles of schooling tend to compartmentalize learning into very discrete entities, such as English in English class or computing in computer science class. In the modern world, regardless of whether the career is scientific or not, skills are generally built from many varied components drawn from a variety of traditional subject domains (Gilbert 2005). This reality is modelled much better by a research approach to learning in science than the traditional transmissive mode evident in high school science.

The capacity for true high school-based astronomical research has exploded over the last twenty years and this has a lot to do with rapid technological advancement. In the early nineties, affordable charge-coupled device (CCD) cameras became available which delivered near-instantaneous images from telescopes rather than the previous long and tedious processing of photographic film (Baruch 1992). While unimaginably slow by present day standards, early modems and bulletin board systems and then the Internet provided a method of 'instantaneous' and cheap long-distance delivery of these images (at least compared to sending the images through the post on disk), as well as the potential for remote control of the observatory itself. 

As it was noted while it was occurring (Baruch 2000, Sadler 2000, Hollow 1998), it is the recent development of fast internet infrastructure that has allowed high school research projects involving astronomy to scale up in recent years. Other Information Technology (IT) developments, such as inexpensive hardware especially in terms of speed and capacity to share large amounts of data, and software to undertake a full analysis, have also contributed. It is also these factors that have allowed remote observing on major research grade observatories to become technically feasible.

Developments in IT even the playing field as less well-resourced schools can freely access such equipment and tools. Prior to this, access may have been the province of the wealthier schools (Gould et al. 2007). Even schools who (due either to budgets, Occupational Health \& Saftery (OH\&S) requirements or the nature of the school's neighborhood) deny students access at night can remotely observe, e.g., from Japan or Australia during a school day in the US. The development of freely available, or relatively inexpensive software, potentially capable of scientific grade measurements, such as Astrometrica (http://www.astrometrica.at/), Makali'i (Horaguchi et al. 2006), ImageJ (Hessman \& Modrow 2008), Aperture Photometry Tool (Laher et al. 2012), and SalsaJ (Doran et al. 2012) go some of the way towards solving earlier issues of inaccessibility to adequate astronomical analysis software (Beare 2006). 

The earliest endeavours were initially made with more modest smaller $\approx12''$ scale telescope systems. As the nineties and early 2000s progressed, larger telescopes and wider, more public, distribution networks grew, e.g., Telescopes in Education (Clark 1998), the Global Telescope Network (http://gtn.sonoma.edu/), the 24$''$ at Yerkes Observatory (Hoette 1998) and the MicroObservatory telescopes (Gould et al. 2006). Recently, larger research grade telescope systems have been constructed which provide telescope time to education projects, e.g., Las Cumbres Observatory Global Telescope (LCOGT; 2x2m Faulkes; Hidas et al. 2008, Gomez \& Gomez 2011; http://lcogt.net) in Hawai'i, USA and Australia, the National Schools Observatory/ Liverpool Telescope (NSO/LT; 1x2m; Steele 2004; http://telescope.livjm.ac.uk/) in La Palma, Canary Islands and MOnitoring NEtwork of Telescopes (MONET) (2x1.2m, http://monet.uni¬goettingen.de) in Texas, US and South Africa. There are also more modest aperture, but still quite sophisticated setups such as the Astronomical Research Institute (http://www.astro-research.org/) in Illinois, US and Tzec Maun Observatory (http://blog.tzecmaun.org/) who host various telescopes in the US and Australia. Notably for optical projects, the ability to connect to a telescope on the other (dark) side of the world opens more possibilities where once it may have required night time observing sessions. As a very useful side-effect, it requires a much decreased learning curve on the part of the teacher to access a remote automated telescope through the internet than to drive a semi-automatic telescope in person (Beare et al. 2003). 

The use of radio astronomy equipment, data and ob¬serving techniques at the high school level has been far less widespread than in optical astronomy. In part, this is due to the greater conceptual difficulty of viewing and interpreting radio data compared with optical data but it is also due to equipment issues. Some schools such as Taunton School in the UK (Hill, 1995) achieved remarkable results in building and using a variety of radio telescopes, including making interferometric observations of the emission from the collision of Comet Shoemaker-Levy 9 with Jupiter. Some schools have used equipment projects such as Radio Jove to establish a radio astronomy provision within their school. Two schools in Australia relocated old dish antennae from the CSIRO Culgoora Radio Heliograph for use within their schools but at least one of these is no longer in use. More recently, several other successful schemes have been implemented in the US and Australia.

In this paper, we aim to provide a summary and analysis of the types of projects that try to, at some reasonable scale, get high school students and/or teachers to undertake, or contribute to, some type of astronomy research. We summarise the criteria used to select projects for this review, define the major dimensions on which the projects differ, provide a description of each project and end with a discussion of the general issues that impact on the success of these projects.

\section{DEFINITION OF AN 'ARiC' PROJECT.}

It is very difficult to categorise these projects into one homogeneous easily inter-comparable group because, while there are many similarities, there are also many differences. The core similarity is that high school students (Yr 9-12) have, at some point, participated in original astronomy using real data from a real instrument, or as close to this as could be realistically practical. 

This is too broad and simple a definition to be generally useful, so we seek to define further criteria for what is, or is not, an Astronomy Research in the Classroom (ARiC) project. Providing and using criteria will always leave out some projects that others may have chosen to leave in. It is, however, beyond the scope of this paper to be too permissive and include every possible astronomy-focussed project. There are other repositories available online that collect lists of programs attempting to bring astronomy data into the classroom (http://nitarp.ipac.caltech.edu/page/other\_epo\_programs).

The following are our nine criteria that a project needs to meet to be defined as an ARiC project. Within reason, these criteria are flexible, especially over the dimension of time; for example, due to technology constraints, earlier projects were much more difficult to run than more recent projects. \\

\hangindent=0.4cm (1) Within the project, there is some capacity for original research.\\

The first criterion is basically a simple definition of an ARiC project. The contribution must be intended to be new, however small, and not be a re-run of previously undertaken research. We are not looking at simulations of real methodologies, such as the CLEA materials  (http://www3.gettysburg.edu/ marschal/clea/CLEAhome.html) 
where the methodological steps that scientists take from observations to results are very well simulated, or canned exercises using real astronomical data contained within existing programs. It can be argued that these activities do have their place in education, and are often vital stepping stones to doing real astronomical research, but they are not the same as actually doing the research. At best, an ARiC project contains within itself mechanisms that can direct the student's novel results towards publication, even in a journal aimed at student publications. \\

\hangindent=0.4cm (2) Data should be from a research-grade instrument and detectors, preferably taken by the students themselves.\\

The second criterion allows multiple astronomical research techniques, perspectives and methodologies to be valid within reason. For instance, optically observing a variable star/s with an amplitude of a large fraction of a magnitude can be undertaken on nearly any clear night, whereas to get a decent colour-magnitude diagram of an open or globular cluster could require precision photometry with outstanding seeing in near-perfect observing conditions. In this case, the direct use of an optical telescope by the student is an option for the first instance above, whereas pulling the data from a research-quality archive is a better option for the second instance. This is a scientific precondition that drives certain pedagogical decisions rather than the other way around. It is important that the end result of a student's research is a scientifically valid contribution, however small, and not, in actuality, a waste of time due to methodological errors. To that end, if the scientific endeavour requires the data-mining of previous observations rather than taking brand new observations, this is also considered to be valid. This criterion also incorporates practical considerations, e.g., if the resources or manpower required to get the project running are prohibitive then requirement for completely new observations is relaxed. This criterion is thus intended to be flexible within reason. \\

\hangindent=0.4cm (3) ARiC projects should focus on the interpretation of data, not just the acquisition. \\

ARiC projects should mimic a real-world research project as much as possible, and encompass at least a few of the steps from the generation of an original idea, through research proposal, literature search, data acquisition, data reduction, and data interpretation, followed by new questions to pursue as a follow-up. Preferably, students should use Flexible Image Transport System (FITS) files, or other standard research data containers and formats, directly. This distinguishes ARiC programs from programs such as the crowd-sourced citizen science programs such as  Galaxy Zoo (http://www.galaxyzoo.org/) or Moon Mappers (http://cosmoquest.org/mappers/moon/) as outlined in Mendez et al. (2010), where many people participate in looking at real data, but  most of the participants typically require a comparatively smaller understanding of the data acquisition, reduction, or interpretation of the ensemble of crowd-sourced results.

Whilst most citizen science projects are not specifically aimed at formal education but rather at informal education, programs such as the Zooniverse are now actively attuned to possibilities and potential of use of programs within the classroom. Such projects are developing useful education tools and educator communities. These are likely to offer some exciting possibilities for ARiC projects in the near future. \\

\hangindent=0.4cm (4) The project must rely fundamentally on the methodologies and typical approaches of the science of astronomy. \\

It is true that there can be much interdisciplinary crossover between astronomy, by which we mean the study of celestial objects, and that of planetary science, whether of Earth or other planets (e.g. http://marsed.asu.edu/mesdt-home, http://minirf.jhuapl.edu/), or that of Space Science, the study of nearby interplanetary space (e.g. http://cse.ssl.berkeley.edu/artemis/epo.html), or with chemistry or biology. For this paper, we choose to focus specifically on astronomical topics. While craters on the Moon and the atmosphere on Mars are interesting extra-planetary topics and are of course not truly distinct from the rest of astronomy (e.g., via extrasolar planets), they often use different language, methodology and approaches to what we typically call 'astronomy'. In the context of this paper, however, since we study asteroids with the general techniques of astrometry and photometry, we make a small exception and include here projects involving asteroids. The study of the nature of the Moon and other planets in our Solar System, however, is beyond the scope of this paper. Unavoidably, this is a criterion with somewhat ill-defined boundaries, boundaries that can also move with time as scientific approaches to the study of natural phenomena themselves change.  \\

\hangindent=0.4cm (5) The focus of the project must be on astronomy rather than general science. \\

There are many organisations and projects around the world that offer generic 'science research projects' where students or educators are generally paired up with a mentor for a one-off science project in any of a wide variety of subjects. The projects we consider here are solely those that are specifically astronomically focussed, and not those that necessarily lead to a one-off science project (see also criterion 7 below). \\

\hangindent=0.4cm (6) Interaction with Students and Teachers must be active, not passive. \\

ARiC projects must involve active, rather than passive, interaction of project personnel with teachers and students. Curriculum repositories such as the SDSS Skyserver materials (http://skyserver.sdss.org/public/en/), the Hands-On Astrophysics/Variable Star Astronomy materials developed by Donna Young, Janet Mattei and John Percy (Mattei et al. 1997, http://www.aavso.org/education/vsa ) or the Chandra activities (http://chandra.harvard.edu/edu/) while valuable, therefore do not qualify. Simple technology provision kits, while enabling science discovery in the classroom, do not necessarily actively promote interaction of scientists with teachers or students. \\

\hangindent=0.4cm (7) Involvement of multiple teachers and/or multiple student groups. \\

We do not focus on single teacher led-projects at specific schools for specific populations of students, even though there have been some fine examples of great projects occurring, such as work at the Taunton Hill Radio Observatory (Hill 1995), Blue Mountains Grammar School (Hollow 2000), The Latin School of Chicago (Gehret et al. 2005). Nor do we consider one-off field trips, since they do not mimic the scientific process. This criterion is to ensure that the ARiC project can work in more than one classroom, involving more than one teacher, enabling at least the possibility that the project is scalable and sustainable. \\

\hangindent=0.4cm (8) Projects are aimed at students or teachers at the high school (School Years 9-12) level. \\

We consider only the high school level components of the projects discussed. Some of the listed projects have, additionally, more substantial undergraduate, elementary or middle school components. To get students interested in science and astronomy, it is well recognised that recruitment needs to begin by the middle school level (Tytler et al. 2008, Barmby et al. 2008, McKinnon \& Geissinger 2002). However, projects that are aimed at this level are of a different nature and are outside the scope of this review. \\

\hangindent=0.4cm (9) Established continuing track record \\
	
In this review, we have omitted some potential projects due to the fact that they are still in development but have mentioned student research as one of their goals as this review is written. We mention these projects here for completeness: Comenius Asteroid Project (http://grudziadz.planetarium.pl/comenius/), Global Jet Watch (http://www.globaljetwatch.net/) or the SOFIA Airborne Observatory (http://sofia.usra.edu/). It is simply our requirement that something substantial must have occurred within the project for us to be able to discuss it. 

There are other projects that have begun but have yet to generate enough data. One example is MONET, using a pair of 1.2m telescopes, one in Texas and one in South Africa (Bischoff et al. 2006), that have had some early success and published some articles (Backhaus et al. 2012, Beuermann et al. 2011 \& Beuermann et al. 2009) but is yet to get rolling in a stable manner. Another example is the Spice-Physics-ICRAR Remote Internet Telescope (SPIRIT) (http://www.spice.wa.edu.au/) pair of observatories at the University of Western Australia. It was officially opened in 2010 for the use of high school students, both remotely and robotically.

\section{DIMENSIONS OF PROJECTS}

Even within the limitations we have imposed, there are nearly as many types of ARiC projects as there are actual projects. They can display different approaches, have widely varying budgets, and hence cannot generally be compared directly. The dimensions outlined here are not rating scales, but rather descriptions of difference. Few of these projects have detailed descriptions in the widely available literature, and so we have defined dimensions in an effort to describe some basic differences and similarities amongst the projects. There are quite a large number of conference proceedings outlining project directions as well as media releases that advertise particular project successes. However, there are many fewer publications of scientific outputs available in the literature (as may be reasonably suspected), and only very rarely are there methodologically strong evaluations of efficacy in terms of student knowledge or motivation outcomes in science. This is also to be expected as a recent analysis of IAU papers (Bretones \& Neto 2011) comes to similar, more rigorous conclusions that research into astronomy education, in general, requires deeper treatments. The lack of good quantitative data in high school astronomy education has been known for a while (Hollow 2000). We now go into detail about the various dimensions we have used to characterise broadly each of the projects. Where each project sits in relation to these dimensions are listed at the top of the description of each of the projects.

\subsection{Teachers / Students / Both}
One of the major dimensions we use below is whether the project focus is on teachers or students as a group or on both groups. Some projects seek to get research undertaken in the classroom via empowering the teacher with inquiry-based scientific skills in the hope that they will in turn get their students to undertake research. In this respect, these projects are also attempting to empower the teacher to better present science to their current and future students in a more engaging manner in addition to implementing authentic research in the classroom.

Other projects act to try to extract particularly interested students from their classrooms as candidates to undertake research. The teacher is used as a recruitment tool to find highly engaged students. Some projects involve the personnel taking control and direction of the classroom activities (either by going into the classroom or bringing the students to the project's staff) leaving the teacher to act in a supervisory or administrative role. 

Teacher professional learning about authentic inquiry-based science education is something that takes a large amount of time, effort and resources (e.g., Supovitz et al. 2000, Gerbaldi 2005, Loucks-Horsley et al. 2003) but can have flow-on benefits for the teachers' future students (e.g., Silverstein et al. 2009), and a multiplicative impact on science education. Student-focussed projects cut straight to the student, eliminating the costs and resources in both money and time, for large amounts of teacher training. 

\subsection{Structured / Guided / Coupled Inquiry}
The tradition of four categories of inquiry were first defined by Schwab (1960) and Herron (1971). Confirmatory Inquiry is where the results of a scientific undertaking are largely known in advance and the scientific endeavour is simply to re-confirm accepted knowledge. By definition, and via criterion number one, this style of inquiry is excluded from this review. Structured Inquiry is where, the larger question and the methodology of an endeavour are quite rigidly set by the teacher/mentor, and the students are led through the methodology to gain a new answer at the end. Guided Inquiry is where, the question is set by the teacher/mentor, but the methodology and process are to a reasonable extent left to the students to design and undertake. Open Inquiry is where students can decide upon their own research question and methodology within the resources available, while the teacher/mentor acts as a guide or facilitator rather than a traditional instructor. 

It is quite unlikely, and pedagogically not feasible, for any student or novice researcher, without prior conceptual and methodological understanding of the scientific field, to launch simply straight into Open Inquiry. More reasonably, the idea of 'Coupled Inquiry' (Dunkhase 2003) is a feasible approach, where students are initially led through either Structured or Guided or some combination to a level where they are capable of undertaking true Open Inquiry. As Open Inquiry has to be preceded by scaffolding at some time and at some level, we leave this out of our discussion. More importantly, scaffolding is required because students cannot do research if they do not understand the content area and methodology of the research field itself (Etkina et al. 2003). 

\subsection{Archival / Pre-Observed / Newly Observed}
This dimension captures whether the project uses data which has been previously analysed and exists in archival datasets, data which has been pre-collected but not analysed or whether the data is collected during the project itself.

\subsection{Selective / Non-Selective / Partially-Selective}
Some projects are open to all who are interested and some are highly academically selective. Others are not necessarily selective, but through their recruitment procedure, end up being selective. 

\subsection{Easily Scalable / Non-Scalable / Scalable with Difficulty}
This dimension is a question of ability to expand given sufficient funding. For instance, an easily scalable project would be one where you would involve twice or more of the population for twice the cost or less. The limiting aspects can relate to the use of a specific instrument or telescope. Unless the telescope is specifically devoted for student use or educational projects, there is an inevitable restriction on time availability if the education projects require their own observations. If the data are primarily archival, the limiting factor may be the number of project personnel recruited to work with the teachers and students. 

\subsection{At School / External / Mixture}
This describes whether the teachers and students undertake the project predominantly in their schools or from home (At School), or at another external institution (External), or it is a mixture (i.e. a workshop at an external institution, then some work back at school or home).

\section{SUMMARY DESCRIPTIONS OF PROJECTS}

In the description of the projects, we have left out the general motivations for the project as they tend to be quite similar. In sum, there is much focus on motivating students in science in high school to inspire them to become scientists or engineers or simply scientifically literate voters in the future through engaging with authentic research experiences. Few science teachers have deep experience in scientific research. This can be a problem because it is difficult to teach what one has never experienced; an analogy would be that one would never hire a sporting-team coach for a high-profile team who had never played that particular game.

There are many statements comparing the gap between "best practice" and "actual practice" in science teaching (e.g., Goodrum et al. 2001; Goodrum, Druhan \& Abbs 2012; European Commission 2007; Osborne \& Dillon 2008) as well as a gap between the pupils', parents' and teachers' perception of the curriculum (Osborne \& Collins 2000). Often there is also reference to the necessity of having enough Science, Technology, Engineering and Mathematics (STEM) trained people or of a particular shortfall within their respective countries (e.g., Select Committee on Science and Technology 2002; European Commission 2004; Tytler 2007; Tytler et al. 2008). Other goals include advancing teacher and student understanding of the nature of science and technology (Duschl, Schweingruber \& Shouse 2007), teacher professional learning (Osborne \& Dillon 2008) and use of information technologies as well as making links between science and other parts of the school curriculum. Rather than focus on individual motivations, we focus on what the projects do or did. We have also steered clear of commenting on the project's applicability to its related curriculum, because this can be highly variable across, as well as within, nations.

It may be noted that the majority of these projects are English-speaking and primarily from the USA with a small number of projects from the UK and Australia. Significant effort was made to find projects in non-English speaking countries, most notably Japan and Continental Europe, however projects fitting the criteria outlined in this paper were not found in our search.

Each of the projects is presented in alphabetical order.  We briefly summarise their major statistics in terms of Project Era, Budget, Scope, and Participants as well as where they fall in terms of the dimensions defined above. The project is then described briefly, with information sourced from a mixture of published papers, conference proceedings and through informal interviews conducted face-to-face, by telephone, via Skype or through email. 

\FloatBarrier

\subsection{Arecibo Remote Command Center (ARCC)}

\FloatBarrier
\begin{table}
\caption{Arecibo Remote Command Center Summary}
\begin{center}
\begin{tabular}{@{}lc@{}}
\hline\hline
%Dimension & Classification accuracy \\
%\hline%
Founded & 2005 \\
End Date  & Continuing \\ 
Budget/Yr  & 10,000USD \\ 
Funding Source  & NSF \& NASA \\ 
Cost to participants & Free \\ 
N (Teachers)  & NA \\ 
N (Students)  & 15-20/year \\ 
N (Schools)  & 5-10/year \\
Location & SW USA \\ 
Website & http://arcc.phys.utb.edu \\ 
\hline\hline
Population  & Students \\ 
Inquiry Depth  & Varies \\ 
Data Rawness  & Varies \\
Selectivity & Selective \\ 
Scalable & No, due to limited telescope time \\ 
On/Off-campus & External \\
\hline\hline
\end{tabular}
\end{center}
\label{tab1}
\end{table}

The Arecibo Remote Command Center (ARCC), based at UT Brownsville, uses the largest single radio dish in the world, the 305m Arecibo Radio Telescope in Puerto Rico, as well as the 64 metre dish at Parkes, NSW Australia (http://www.parkes.atnf.csiro.au/) and the Green Bank Telescope (GBT; https://science.nrao.edu/facilities/gbt/). The instruments are used to get high school and undergraduate students to make observations and undertake analyses in searching the galaxy for pulsars. There is also an ARCC at the University of Wisconsin, Milwaukee (http://www.gravity.phys.uwm.edu/arcc/). High School students can access a summer school program "21st Century Astronomy Ambassadors", which involves a three- week residential program and further ongoing involvement. Students use archival data from GBT, Arecibo, Parkes and LWA. There is also capacity to conduct new observations with GBT and Arecibo as well. Students have discovered 25 pulsars; in 2011 they discovered approximately 40\% of all pulsars discovered that year.

\FloatBarrier
\subsection{Bradford Robotic Telescope (BRT)}

\FloatBarrier
\begin{table}
\caption{Bradford Robotic Telescope Summary}
\begin{center}
\begin{tabular}{@{}lc@{}}
\hline\hline
%Dimension & Classification accuracy \\
%\hline%
Founded & 1993 \\
End Date  & Continuing \\ 
Budget/Yr  & 280,000UKP \\ 
Funding Source  & Self-funded \\& + 1 University Position \\ 
Cost to participants & 70UKP/year/school \\ 
N (Teachers)  & 4000 Total, 450 Per Year \\ 
N (Students)  & 90,000 Total, 13,000 Per Year \\ 
N (Schools)  & 1500 Total, 200 per Year \\
Location & UK, Europe \\ 
Website & http://telescope.org \\ 
\hline\hline
Population  & Both \\ 
Inquiry Depth  & Structured \\ 
Data Rawness  & Newly Observed \\
Selectivity & Non-selective\\ 
Scalable & Easily Scalable \\ 
On/Off-campus & At School \\
\hline\hline
\end{tabular}
\end{center}
\label{tab1}
\end{table}

Originally commissioned in 1993 in the Yorkshire Pennines (Baruch 2000), the Bradford Robotic Telescope (BRT) is now a 0.365m completely autonomous robotic telescope sited in Mount Teide, Tenerife in the Canary Islands. Initially, the telescope was begun as an engineering/astronomy project, but over time transitioned into a primarily educational/outreach organisation, as teacher interest was unexpectedly high and sustained and education related funding started to flow towards the project. BRT allows schools and the general public to submit observing 'jobs' to the telescope over the internet through a portal website. The images are then queued, observed robotically, and then returned to the user. The telescope itself has three main cameras, a 'constellation' camera (40 degree field of view (FOV)), a 'cluster' camera (4 degree FOV) and a 'galaxy' camera (20 arcminute FOV). Requested images are usually obtained and returned within a night or two, depending on weather conditions and celestial location of the object.

Schools are charged for usage as the administrating university wants the project to pay its own way. While amateur astronomers, some of whom looked for supernovae and NEOs, have had free use for many years, they are also now charged. Historically, the telescope was run mainly through a variety of grants and subsidised by in-kind support from the university. Nowadays, BRT only receives grants for educational research and outreach work. There are six staff members, five of whom are covered by the income from the project, while the PI's income comes from university. Overall, there are 100,000s of school users (and 30,000 amateur users). 

The BRT is focussed at Grades 3 to 6 (Key Stage 2) up to Grades 10 and 11 (Key Stage 4) and provides education and teaching materials for all years. The main focus of the project is on education and learning rather than on authentic research itself. There are possibilities for open-ended use of the BRT through its various programs, although currently the inability of the scheduler to acquire calibration frames easily or series of related images limits its use to general colour imaging. While it is possible for the students to undertake research through the BRT, the general focus is on getting all students familiar with the night sky and pushing the more interested students to larger research-based projects such as others listed in this paper. 

Largely, teachers have approached BRT to get involved themselves through finding the project through the web or via word of mouth. BRT also actively looks for teachers by choosing an area, looking for supportive grants in the area and then linking up with the secondary schools and the primary schools that feed the secondary schools. A large recent thrust of the BRT project is to provide professional development for teachers in the area of astronomy subject knowledge because it is perceived that one source of what discourages students in science is the teacher's relatively poor perception of science. The program provides professional development (PD) for a teacher in his or her classroom. This has resulted in a far larger usage of the telescope than other more traditional models in terms of number of students involved, projects undertaken, and activity on the website. In one year, the project generally provides in-classroom PD for 200 to 300 teachers.

\FloatBarrier
\subsection{Charles Sturt University Remote Telescope Project}

\FloatBarrier
\begin{table}
\caption{Charles Sturt University Remote Telescope Summary}
\begin{center}
\begin{tabular}{@{}lc@{}}
\hline\hline
%Dimension & Classification accuracy \\
%\hline%
Founded & 1999 \\
End Date  & Continuing \\ 
Budget/Yr  & 2500AUD \\ 
Funding Source  & Self-funded \\ 
Cost to participants & Donations \\ 
N (Teachers)  & 70/Year, 800 Total \\ 
N (Students)  & 1500/Year, 25,000 Total \\ 
N (Schools)  & 50/Year, 400 Total \\
Location & Australia, Europe,\\& Canada, USA \\ 
Website & http://www.csu.edu.au/\\&telescope/ \\ 
\hline\hline
Population  & Both \\ 
Inquiry Depth  & Structured \\ 
Data Rawness  & Newly Observed \\
Selectivity & Non-selective \\ 
Scalable & Scalable \\ 
On/Off-campus & At School \\
\hline\hline
\end{tabular}
\end{center}
\label{tab1}
\end{table}

The Charles Sturt University Remote Telescope Project (CSURTP) was initially set up, in 1999, for use specifically by upper primary school (Years 5 \& 6) students (McKinnon \& Mainwaring 2000) as a 12$''$ telescope that could be used to gather images in real time via the internet. This has since been expanded to include high school level (Years 7-10) (McKinnon 2005, Danaia 2006) and beyond. Some senior students in Australia and Canada have used the system to acquire data on variable stars as part of their science assessment requirements to undertake an individual research projects.

The primary use of the CSURTP is driven through the 'A Journey through Space and Time' curriculum materials. These materials lead teachers and their students through preparing and planning for using the telescope remotely as well as interpreting their observations after the observing session. Many different sets of materials are provided that allow students to explore various aspects of astronomy from simple astrophotography up to monitoring variable stars and creating colour-magnitude diagrams of stellar clusters. The depth to which the teacher goes is decided by the teacher in their particular context.  

The 'Eye Observatory' Project, using this facility and materials developed for the CSURTP, provided an educational evaluation of over 2000 students and 101 teachers from 30 schools as they went through an investigative structured inquiry procedure (Danaia 2006). Highly significant gains in terms of the students' understanding of astronomical phenomena and a marked reduction in alternative conceptions via using a subset of questions from the Astronomy Diagnostic Test (Hufnagel et al. 2000) were achieved. Students' perceptions of science in school and in the larger world showed significant gains, although these varied dramatically between differing schools, teachers and student groups. 
 
\FloatBarrier
\subsection{Faulkes Telescope Project}

\FloatBarrier
\begin{table}
\caption{Faulkes Telescope Project Summary}
\begin{center}
\begin{tabular}{@{}lc@{}}
\hline\hline
%Dimension & Classification accuracy \\
%\hline%
Founded & 2003 \\
End Date  & Continuing \\ 
Budget/Yr  & Internal University \\ 
Funding Source  & Initially Grant-funded, \\
& now internal university.\\ 
Cost to participants & Free \\ 
N (Teachers)  & Exact Numbers Unknown \\ 
N (Students)  & Exact Numbers Unknown \\ 
N (Schools)  & 200-250 Schools per Year \\
Location & UK, Europe \\ 
Website & www.faulkes-telescope.com \\ 
\hline\hline
Population  & Teachers \\ 
Inquiry Depth  & Structured/Guided \\ 
Data Rawness  & Newly Observed \\
Selectivity & Not Selective \\ 
Scalable & Scalable \\ 
On/Off-campus & At School \\
\hline\hline
\end{tabular}
\end{center}
\label{tab1}
\end{table}

The Faulkes Telescope Project (FTP) is hosted at the University of Glamorgan. It was initially based in Cardiff, Wales from 2003 to 2010, using Faulkes Telescope North (FTN) from early 2004 and Faulkes Telescope South (FTS) from early 2006. The FTP focuses on providing access to telescope time to UK schools and educational groups for general use, as well as materials facilitating student research (Lewis et al. 2010, Roche \& Szymanek 2005) through collaboration with FTP astronomers. 

FTP was initially set up as an educational team by the Dill Faulkes Educational Trust (http://www.faulkes.com/dfet/) who had constructed the Faulkes Telescopes via a substantial GBP10 million donation in the late 1990s/early 2000s. In 2003, just before FTN came online, the team started building an education program using funds provided by Dr Dill Faulkes and also a 600,000GBP grant from the UK government. In particular, this program focussed on teacher training to get teachers capable of using the telescopes. It was, from late 2005 to 2010, an educational and research arm of LCOGT, but is now an internally funded university program.

Initially, many of the schools that joined were from the private sector in the UK rather than the publicly funded sector. The reason for this seemed to be because teachers from these schools are usually more highly qualified and have greater control over the curriculum offerings at their schools. Over time however, publicly funded schools slowly came on board. There are 200-250 schools using the telescope regularly out of about 4500 high schools in the UK. The project is currently working to branch out into Europe; they have some European schools on board in Portugal, France, and Germany, amongst other countries. With the new European focus, there are more projects coming on board as well as well as involvement with ESA. 

The project is trying to build projects where astronomy researchers require data and teachers and students can be involved in the acquisition of those data using the Faulkes Telescopes. While there were some issues with quality control of the data, the early feedback indicated that schools liked the fact that they were helping real astronomers. The projects had to be something that students could actually get their heads around, rather than an obscure topic where they may not know what they are doing, and preferably produce something aesthetically pleasing. Topics that seemed to work well were topics such as open clusters as well as galaxies, but the asteroid and comet studies were what really had an impact and which have also seen significant involvement by the amateur community. Approximately half of the schools involved undertake just "pretty picture imaging" once a year during their astronomy topic. Not all schools are doing astronomy as part of the formal curriculum; some are using FTP materials in their after-school or lunch-time programs. The other half regularly engage with projects that have been made available online.

A recently tested approach has been to suggest targets for the students to observe for scientists during their ordinary observing sessions. The most popular though, in terms of getting schools on board, was advertising "themed observing days/weeks/months", where an entire period of time (usually a day) is spent looking at a particular object, such as an asteroid or a group of galaxies. Schools had to sign up to be allocated time, usually 30 minutes, on that day and these were generally two to three times oversubscribed. In return, scientists got eight hours of concentrated data.   

A group of UK schools helped name an asteroid, 2004WB10 'Snowdonia' that they observed with FTN as well as 2005HJ4 'Haleakala'. Working with Richard Miles, other schools  determined the rotation rate of the rapidly rotating asteroid 2008HJ (Lewis \& Roche 2009). In addition, there are many asteroid observations published through the Minor Planet Centre (http://www.minorplanetcenter.net/). Students are also involved in collecting data for an X-ray Binary monitoring project (Lewis et al. 2008), although they do not analyse the data directly due to the restrictive software requirements.  

\FloatBarrier
\subsection{Goldstone Apple Valley Radio Telescope (GAVRT)}

\FloatBarrier
\begin{table}
\caption{Goldstone Apple Valley Radio Telescope Summary}
\begin{center}
\begin{tabular}{@{}lc@{}}
\hline\hline
%Dimension & Classification accuracy \\
%\hline%
Founded & 1997 \\
End Date  & Continuing \\ 
Budget/Yr  & Unknown \\ 
Funding Source  & Congressional appropriations \\ 
Cost to participants & Free \\ 
N (Teachers)  & 490 Total \\ 
N (Students)  & 32,000 Total \\ 
N (Schools)  & 290 Total \\
Location & USA and 13 other countries \\ 
Website & www.lewiscenter.org/gavrt \\ 
\hline\hline
Population  & Teachers \\ 
Inquiry Depth  & Structured \\ 
Data Rawness  & Newly Observed \\
Selectivity & Non-selective \\ 
Scalable & Scalable \\ 
On/Off-campus & Mixture \\
\hline\hline
\end{tabular}
\end{center}
\label{tab1}
\end{table}

The GAVRT Program was initially run through JPL and the Lewis Center for Educational Research. The core of the GAVRT is a 34-meter radio telescope previously used as a node in NASA's Deep Space Network, which is now provided for teachers and students to use. As of early 2012, GAVRT had served 490 Teachers in 290 Schools, mainly in the US, but also 14 other countries including Chile, Germany, Italy, Japan, South Korea and the UK. Current statistics can be seen here (http://gsc.lewiscenter.org/gavrt/schools.php). Permanent staff in the project include a part-time manager and a full-time and part-time operator.

The main professional learning approach for teachers is via a 5-day workshop providing skills in radio astronomy basics, radio telescope control and curriculum support. Project scientists take time out of their schedules to visit students in their classroom or connect via video conferencing as well as answering questions by email. The current astronomical focus of GAVRT is on Jupiter, Quasars and the Search for Extraterrestrial Intelligence (SETI). The students have their observing routine relatively well laid out for them in the curriculum materials in terms of what scans to run, frequencies, sources and rates, although in the SETI program there is more freedom in terms of target selection and analysis. Access to data is through an online access tool, but during the sessions, students undertake calculations to convert the data into something they can understand.  	

\FloatBarrier
\subsection{Hands on Universe}

\FloatBarrier
\begin{table}
\caption{Hands On Universe Summary}
\begin{center}
\begin{tabular}{@{}lc@{}}
\hline\hline
%Dimension & Classification accuracy \\
%\hline%
Founded & early 1990s \\
End Date  & early 2000s \\ 
Budget/Yr  & 400,000USD \\ 
Funding Source  & Initially NSF \\ 
Cost to participants & Free \\ 
N (Teachers)  & 800 Total \\ 
N (Students)  & Unknown \\ 
N (Schools)  & Unknown \\
Location & Global \\& but predominantly USA \\ 
Website & http://www.globalhou.net \\ 
\hline\hline
Population  & Both \\ 
Inquiry Depth  & Structured \\ 
Data Rawness  & Varies \\
Selectivity & Not Selective \\ 
Scalable & Scalable \\ 
On/Off-campus & At School \\
\hline\hline
\end{tabular}
\end{center}
\label{tab1}
\end{table}

Hands on Universe (HOU) (http://www.handsonuniverse.org/) started in the early 1990s as a project to get students, primarily in grades 9-12 (Rockman 1994) actively 'hands-on' with real noisy astronomical data, rather than traditional lecture-style teaching. The current version of the project is run through Global Hands-on Universe (GHOU, http://www.globalhou.net/) at UC Berkeley in the USA and Nuclio in Portugal, and is primarily a teacher training project rather than an ARiC style project.

The earlier form (pre-2000) of the HOU project was supported by much more substantial funding than the current form. A lot of this early funding went into development on various fronts such as software, telescope infrastructure, curriculum development, and teacher training. During this earlier era, HoU attempted to create a network of observatories from various university, private, and public institutions who could provide a percentage of their observing time to the HoU project. This network initially started out using the 30$''$ automated telescope at Leuschner Observatory at U.C. Berkeley (Asbell-Clarke et al. 1996) and expanded to a variety of telescopes around the world (Boer et al. 2001).

The materials originally created for Hands-On Universe are now part of the Global Systems Science curriculum (http://www.globalsystemsscience.org/). Teacher training in the earlier version of HOU was provided during summer workshops using a model where teachers who had previously undertaken the workshop trained the next cohort of teachers and acted as 'Teacher Resource Agents'. By mid-1998, it was estimated there were over 500 US teachers and hundreds of international teachers who had undertaken the training under a TRA (Pennypacker 1998). Projects included measuring variable stars, sizes of objects in the Solar System and telescope construction (Pennypacker \& Asbell-Clarke 1996), as well as more open-ended curricula via acquiring images from the automated telescopes. Custom HOU software that performed a variety of image display and analysis tools was constructed for this purpose, but has since been dropped as more recent technologies and better software has arisen in the community.

A web based asteroid search project using archive images from the 4m CTIO Blanco telescope was designed in the mid 1990s. Using an image subtraction method, many main belt asteroids were found, but the most interesting success was that students had found a Trans-Neptunian object (1998 FS144) which was reported to the Minor Planet Centre (MPC) (Pack 2000). Two students from Oil City High School involved with the project also helped measure the lightcurve of a supernovae and were co-authors on a scientific paper (Richmond et al. 1996).

\FloatBarrier
\subsection{Hawaii' Student Teacher Astronomy Research Program (HISTAR)}

\FloatBarrier
\begin{table}
\caption{HISTAR Summary}
\begin{center}
\begin{tabular}{@{}lc@{}}
\hline\hline
%Dimension & Classification accuracy \\
%\hline%
Founded & 2007 \\
End Date  & Continuing \\ 
Budget/Yr  & 15,000USD \\ 
Funding Source  & Grants, Donations,\\& Agency Funds \\ 
Cost to participants & 150USD \\ 
N (Teachers)  & Unknown \\ 
N (Students)  & 12-16/year \\ 
N (Schools)  & NA \\
Location & Hawaii', Continental US \\ 
Website & http://www.ifa.hawaii.edu/ \\
& UHNAI/HISTAR.html \\ 
\hline\hline
Population  & Both \\ 
Inquiry Depth  & Guided \\ 
Data Rawness  & Newly Observed \\
Selectivity & Selective \\ 
Scalable & Scalable \\ 
On/Off-campus & External \\
\hline\hline
\end{tabular}
\end{center}
\label{tab1}
\end{table}

The Hawai'i Student/Teacher Astronomy Research Program (http://www.ifa.hawaii.edu/UHNAI/HISTAR.html) started in 2007. Approximately seventeen 12 to 16 year old students per year are selected from local Hawaiian schools, as well as teachers, to attend the week long 9am to 9pm workshop at the Institute for Astronomy at the University of Hawai'i. During the workshop accompanied by traditional lectures, students undertake a research project with an astronomical mentor lasting the full week. They do a 10-minute presentation of their results at the end of the program. They also have observation sessions with the DeKalb 16$''$ Telescope in Indiana (http://www.starkey.ws/new observatory.html) and the 2-meter Faulkes Telescope in Hawai'i (lcogt.net), and attend a variety of field trips to planetariums and university departments.   

The program authors' intention is to equip students and teachers with the background knowledge to undertake astronomy research that they will continue on return to their schools with the help of a mentor. Projects include such topics as variable star photometry, comets, extra-solar planets, and rotational velocities and masses of galaxies (Garland et al. 2008). About 30\% of each year's intake are returning HI STAR students who have completed a research project and have come back to learn more. These students have entered their research in science fairs and won a variety of awards (Kadooka 2011). This project is sponsored by NASA's National Astrobiology Institute (NAI), Institute of Astronomy (IfA) and the Center for Computational Heliophysics in Hawai'i (C2H2) as well as contributions from a NASA IDEAS grant and LCOGT. 

\FloatBarrier
\subsection{International Asteroid Search Campaign (IASC)}

\FloatBarrier
\begin{table}
\caption{International Asteroid Search Campaign Summary}
\begin{center}
\begin{tabular}{@{}lc@{}}
\hline\hline
%Dimension & Classification accuracy \\
%\hline%
Founded & 2006 \\
End Date  & Continuing \\ 
Budget/Yr  & None until 2012 \\ 
Funding Source  & Small grants \\ 
Cost to participants & Free \\ 
N (Teachers)  & 500/Year \\ 
N (Students)  & 5000/year \\ 
N (Schools)  & 500/year \\
Location & Global \\ 
Website & http://iasc.hsutx.edu \\ 
\hline\hline
Population  & Teachers \\ 
Inquiry Depth  & Structured \\ 
Data Rawness  & Pre-observed \\
Selectivity & Non-selective \\ 
Scalable & Scalable \\ 
On/Off-campus & At School \\
\hline\hline
\end{tabular}
\end{center}
\label{tab1}
\end{table}

Commencing in October 2006, the focus of the Hardin-Simmons University-based International Asteroid Search Campaign (IASC) is to guide students through the process of astrometrically analysing supplied asteroid and NEO observation data sets to, at the final stage, publish their discoveries and measurements to the Minor Planet Center (Miller et al. 2008). The data were initially collected solely for this purpose using two (0.81m and 0.61m) telescopes at the Astronomical Research Institute in Illinois (http://www.astroresearch.org/). Now data collection has progressed to include the Pan-STARRS telescope, the 2m Faulkes Telescope in Hawai'i, a 1.3m instrument at Kitt Peak and the 60-90cm Schmidt Telescope from National Astronomical Observatories of China, as well as a variety of other telescopes and ones coming online. 

There is no cost to participate. The program ran with help from volunteers and without funding until 2012 when grant funding started. Funding included a grant from the European Commission through the Paris Observatory and supplemental funding from Texas. IASC is a volunteer organisation; the data reduction team consists of four people around the world who donate their time. They validate the MPC reports sent in by students after which they send the validations to the MPC. Teachers are predominantly self-trained using supplied materials, practice data sets and the Astrometrica software package (http://www.astrometrica.at/). Online help is provided at any time, although there have been the occasional face-to-face training session as well. The project plans to offer more face-to-face sessions. The training for the main belt observations must be done and class implementation must occur before teachers are invited back to undertake further training using the raw data used in the NEO campaigns. The actual structure of any in-class implementation is left heavily up to the teacher to design.

Prior to delivery to the teacher, the data are pre-screened for astrometric quality. IASC is performed on pre-set 30-day main belt and 60-day NEO asteroid campaigns each year, with team members post-validating any discovery made by the students. Approximately 500 schools in 60 countries are now participating. There are ten five-week segments and anywhere from 30 to 80 schools participating in any one segment. Participation is largely dependent on the telescope resources available during the period. The ultimate goal and capacity is to grow to around 1000 schools. To achieve this, IASC is looking for grant funding to automate a lot of the processing and MPC reports. If professional staff can be hired, then the project could be expanded to 5,000-10,000 schools.

As of 2008, with the publication of their outline paper cited above, 36 new Main Belt asteroids, 197 NEOs and one comet were discovered (Miller et al. 2008). As of this writing, the number is around 350 asteroids, of which 15 are now numbered and being named by their students. The outputs include more than 100 Minor Planet Center Circulars published, two Earth-threatening NEOS and a Trojan asteroid of Jupiter (Miller, 2012 private communication). 

\FloatBarrier
\subsection{MicroObservatory / Other Earths, Other Worlds}

\FloatBarrier
\begin{table}
\caption{MicroObservatory Summary}
\begin{center}
\begin{tabular}{@{}lc@{}}
\hline\hline
%Dimension & Classification accuracy \\
%\hline%
Founded & 1988, first light 1995 \\
End Date  & Continuing \\ 
Budget/Yr  & Unknown \\ 
Funding Source  & NSF \\ 
Cost to participants & Free \\ 
N (Teachers)  & Unknown \\ 
N (Students)  & Unknown \\ 
N (Schools)  & Unknown \\
Location & Global \\ 
Website & mo-www.cfa.harvard.edu/ \\
& MicroObservatory/
 \\ 
\hline\hline
Population  & Both \\ 
Inquiry Depth  & Structured \\ 
Data Rawness  & Newly Observed \\
Selectivity & Non-selective \\ 
Scalable & Scalable \\ 
On/Off-campus & At School \\
\hline\hline
\end{tabular}
\end{center}
\label{tab1}
\end{table}

MicroObservatory saw first light in 1995 (Sadler et al. 2000) (http://mo-www.harvard.edu/MicroObservatory/). It is a series of five 6-inch automated waterproof telescopes built by the Harvard-Smithsonian Center for Astrophysics (CfA) at 2 locations: the CfA itself and Whipple Observatory in Arizona. It provides images freely with a specific focus on education. Participation in the project is free of charge to all schools and is primarily funded by the National Science Foundation (NSF). 

The earlier MicroObservatory project has since evolved into 'Laboratory for the Study of Exoplanets' via a pilot project called 'Exploring Frontiers of Science with Online Telescopes' (Gould et al. 2012). The focus of both projects has been to engage students in detecting and describing exoplanets using the MicroObservatory telescopes. The focus is on providing online tools to explore exoplanets, albeit exoplanets that have previously been identified by astronomers. The students take the data themselves.
 
The telescopes are operated robotically. However, if different schools request the same object with the same exposure time and filter, the telescopes will only take one, rather than multiple, sets of data. Around 20 known exoplanets are large enough (and their host stars bright enough) to have transits detected by the MicroObservatory telescopes. By 2006 (Gould et al. 2007), more than half a million 650x500 pixel, 12 bit images covering one square had been taken primarily for students, but also for public outreach endeavours (Sadler et al. 2001). The telescopes can be used in three modes: 'Full Control' where the user can drive the telescope remotely in real time, 'Guest Observer Mode', where users request images from the telescopes in a more traditional robotic fashion, and 'Research Mode', where Target of Opportunity (TOO) triggers can take over the telescope to observe time-dependent transient phenomena.

Personnel at the MicroObservatory have created their own simple image processing program to deal with the FITS files from the telescopes, with the most complexity lying in the creation of colour images from the raw FITS files and making simple animations. Curriculum materials are also provided that lead the students through the basic concepts of telescopes and what they can see so that they can undertake simple structured inquiry with the telescopes.  

\FloatBarrier
\subsection{NASA/IPAC Teacher Archive Research Program (NITARP) / Spitzer Space Telescope Research Program}

\FloatBarrier
\begin{table}
\caption{NITARP Summary}
\begin{center}
\begin{tabular}{@{}lc@{}}
\hline\hline
%Dimension & Classification accuracy \\
%\hline%
Founded & 2005 \\
End Date  & Continuing \\ 
Budget/Yr  & 300,000USD \\ 
Funding Source  & NASA \\ 
Cost to participants & Free \\ 
N (Teachers)  & 18/Year, 87 Total \\ 
N (Students)  & > 200 Total \\ 
N (Schools)  & 18/Year, 85 Total \\
Location & USA \\ 
Website & http://nitarp.ipac.caltech.edu \\ 
\hline\hline
Population  & Teachers \\ 
Inquiry Depth  & Guided/Coupled \\ 
Data Rawness  & Archival and Pre-Observed \\
Selectivity & Selective \\ 
Scalable & Scalable \\ 
On/Off-campus & Mixture \\
\hline\hline
\end{tabular}
\end{center}
\label{tab1}
\end{table}

The Spitzer Space Telescope Research Program ran from 2005 to 2009. It transitioned into the NASA/IPAC Teacher Archive Research Program (NITARP), running from 2010 until the present (http://nitarp.ipac.caltech.edu). The original program,  run by the Spitzer Science Center (SSC) located at the California Institute of Technology in Pasadena, CA, USA, which is part of the Infrared Processing and Analysis Center (IPAC). The discretionary funding currently comes from NASA, and until 2014 some EPO funds tied to the archives housed at IPAC. In 2014, the funding was reduced from \$300K to \$200K meaning a reduction in teachers per year from 18 to 9.

NITARP partners small groups of educators with a mentor professional astronomer for an original year-long research project. It involves three trips for participants (and two trips for two of each of their students), all of which are paid for by the program. Participants kick off their year by attending an American Astronomical Society (AAS) winter meeting, then come with $\lesssim$2 students per educator to visit Caltech for a week in the summer, and then return to an AAS meeting (with $\lesssim$2 students per educator) to present their results. Much of the work is done remotely during the rest of the year. Each team consists of a mentor astronomer, a mentor teacher (who has been through the program before), and typically 2-4 new teachers.

An annual program cycle runs for a calendar year with applications available annually in May and due in September. Teachers are encouraged to bring students on the second two trips. The number of teams that NITARP can support in any given year depends primarily on having an available mentor scientist and enough money to support the trips.

NITARP and its predecessor have historically been aimed at high school classroom teachers, but now solicits applications from teachers in 8th grade through to community college, and from both formal and informal educators (e.g., museum staff). Applicants must have had some experience in astronomy, preferably having worked with astronomy data but not necessarily having conducted any research. In the earlier Spitzer Space Telescope version of the project, teachers must have been graduates of the NOAO Research Based Science Education (RBSE) program.

NITARP teams all must use astronomical data housed at IPAC. Thus, the NITARP teams use the same data in the form used by professional astronomers. All of these data are necessarily pre-observed, but they are reduced to varying degrees. In the early (Spitzer) years of the program, participants in the program were granted relatively small amounts of telescope time on the Spitzer Space Telescope to conduct their research.

Some limited evaluation was conducted as part of the Spitzer program before NITARP started. At that point, 32 teachers had been through the program and had brought a total of 79 students to either Caltech or the AAS, and over 1200 students had used Spitzer data in their classrooms. The teachers and their students made nearly 200 presentations reaching an estimated 14,000 people. A little over 100 students reported that the program had influenced them to pursue careers in science, and 42 had entered Spitzer-related science-fair projects, including contributions that made it to the Intel Science Fair level. A detailed evaluation is being conducted on the NITARP 2013 participants with results expected in 2014, with results expected in 2014. A brief survey was also conducted on the entire set of NITARP+Spitzer alumni in Spring 2014, with 50\% survey response rate. As reported in Rebull et al. (2014), the NITARP participants estimate that ~13,000 students have benefitted from skills or resources the educator learned about via NITARP, ~21,000 students are taught by NITARP educators per year, and ~4300 other educators have been reached with NITARP information.

There have been 43 science and 49 education AAS posters presented by NITARP-affiliated educators and scientists.  There have been 5 refereed journal articles in major professional astronomy research journals that directly involve NITARP teachers and scientists (Howell et al., 2006,  Guieu et al., 2010, Howell et al., 2008, Rebull et al., 2011, Rebull et al. 2013). There have been 2 more astronomy journal articles involving NITARP scientists, describing the software developed in conjunction with NITARP and its Spitzer predecessor (Laher et al. 2012). Finally, there is one more refereed article written by a NITARP alumni teacher for The Physics Teacher (Pereira et al. 2013).  

\FloatBarrier
\subsection{Pisgah Astronomical Research Institute}

\FloatBarrier
\begin{table}
\caption{Pisgah Astronomical Research Institute Summary}
\begin{center}
\begin{tabular}{@{}lc@{}}
\hline\hline
%Dimension & Classification accuracy \\
%\hline%
Founded & 1999 \\
End Date  & Continuing \\ 
Budget/Yr  & 5,000USD \\ 
Funding Source  & Private Donations \\ 
Cost to participants & \$100 \\ 
N (Teachers)  & 300 Total, 15/Year \\ 
N (Students)  & 400 Total, 45/Year \\ 
N (Schools)  & 500 Total, 30/Year \\
Location & North Carolina,\\& Southern USA\\ 
Website & http://www.pari.edu/ \\ 
\hline\hline
Population  & Students \\ 
Inquiry Depth  & Structured/Guided \\ 
Data Rawness  & Newly Observed \\
Selectivity & Semi-Selective \\ 
Scalable & Not easily Scalable  \\ 
On/Off-campus & External \\
\hline\hline
\end{tabular}
\end{center}
\label{tab1}
\end{table}

The site upon which the Pisgah Astronomical Research Institute (PARI) stands is about 50 years old.  The site was originally used for line-of-sight communication and telemetry with satellites as well as for human spaceflight. As the Apollo program wound down, the site was transferred to the U.S.  Department of Defense (DoD) in the early 80s and turned it into a military installation until 1995 when it was shut down. In 1999 the facility was purchased from the DoD and is now used as a science and technology center supporting public, private, commercial and government endeavours.

PARI's mission is primarily educational; although initially aimed at university level, it was discovered that there were many opportunities at different year levels. PARI also provides school group tours and group visits that occur daily.  The instruments on the site used for educational programs include a variety of optical telescopes (10-inch to 14-inch) for both imaging and spectroscopy and 'Smiley', the 4.6m radio telescope (PARI, 2011). There are two main PARI high school programs; the Space Science Lab Program (SSLP), which is funded through a Burroughs Wellcome Fund Grant, and the Duke Talent Identification Program (Duke TIP), which is run from Duke University.

The Duke TIP program is competitive entry, takes in 30 students/year and generally has a waiting list. In the SSLP, while also intended to be competitive, the students are more directly recruited by PARI staff giving talks in schools at the invitation of teachers. The drawing area for SSLP is in Western North Carolina, in general a low socioeconomic area, whereas Duke TIP is nationwide, although draws more frequently from the Southern states of the US. 

In the DUKE TIP program, the students have to make posters on research they perform at the site, which are printed and presented on the PARI website. More directed research is possible; however, the nature of the research is restricted by the 2 week limit. Students are helped in brainstorming ideas to come up with possible projects plausible within the constraints of time and instrumentation. Projects vary from spectroscopic binary stars, to the rotation of Saturn's rings, to radio jets in galaxies, amongst many others, all having some substantial component using a real telescope.

The SSLP uses Radio Jove as an introduction to radio astronomy and the technical side of instrumentation. They then use the Smiley telescope upon which they can book observing time. Usually it is used primarily to make continuum observations of the Sun, which are comparable to the Jovian system. There are 500 users in total registered on Smiley; the majority of whom are teachers who use it once per year, while approximately 10-20 per year use it for a directed research project per year. 

\FloatBarrier
\subsection{Pulsar Search Collaboratory (PSC)}

\FloatBarrier
\begin{table}
\caption{Pulsar Search Collaboratory Summary}
\begin{center}
\begin{tabular}{@{}lc@{}}
\hline\hline
%Dimension & Classification accuracy \\
%\hline%
Founded & 2008 \\
End Date  & Continuing \\ 
Budget/Yr  & 900,000USD \\ 
Funding Source  & Initially NSF, now volunteer \\ 
Cost to participants & Free \\ 
N (Teachers)  & 106 Total \\ 
N (Students)  & 2431 Total \\ 
N (Schools)  & 103 Total \\
Location & West Virginia and surrounds \\ 
Website & pulsarsearchcollaboratory.org \\ 
\hline\hline
Population  & Both \\ 
Inquiry Depth  & Guided \\ 
Data Rawness  & Pre-observed \\
Selectivity & Not Selective \\ 
Scalable & Scalable \\ 
On/Off-campus & Mixture \\
\hline\hline
\end{tabular}
\end{center}
\label{tab1}
\end{table}

Beginning in 2008, the Pulsar Search Collaboratory (PSC) was a 3 year NSF-funded project jointly run by the National Radio Astronomy Observatory (NRAO) and West Virginia University (Rosen et al. 2010) which is currently continuing on a volunteer-run basis. While undergoing maintenance, 1500 hours of observing data are recorded via drift scanning at the GBT. Of these, 300 hours were reserved for the use of high school students who were initially from West Virginia Schools, but over time, this expanded to surrounding states. The total funding for the three-year project was around \$900,000, which predominantly funded participants' food and housing, and  salaries for those involved. 

The PSC runs with an approach where teacher and student leaders are trained in radio and pulsar astronomy techniques at a three-week workshop held annually. They go back to their schools and recruit other interested teachers and students. Admission is not selective, but tends to attract students at the top of their classes.

The students attend the last six days of the workshop while the teachers are there for the full three weeks. A 40 foot telescope is used to become familiar with telescope usage and concepts before they are moved onto a pulsar project using the GBT. The teachers and students go back to their schools to work on the project over the course of a year, whether in class or in an astronomy club, with the participating students acting as team leaders. At the end of the year, PSC organises a seminar at West Virginia University for students and teachers to present their results as well as also providing videoconferencing to any follow-up observing sessions based on student analysis. 

Evaluation reports suggest that interest is significantly increased in STEM careers after participation in PSC, with 39\% reporting increased interest and 55\% reporting similar interest after participation. They also see a significant increase in self-confidence and self-efficacy in science, particularly amongst girls (Rosen et al. 2010). A large amount of the gain in engagement comes from students going to Green Bank or West Virginia University to meet up with other schools or other teams. There have been a number of discoveries by students of new pulsars and one potential rotating radio transient  (https://sites.google.com/a/pulsarsearchcollaboratory.com/pulsar-search-collaboratory/Home/new-psc-pulsars). One of the students, the discoverer of a potential rotating radio transient, Lucas Bolyard, got to meet the President and First Lady at a White House star party. A scientific paper outlining the discovery and timing of five pulsars identified in the project with many student co-authors has been published (Rosen et al. 2013)

\FloatBarrier
\subsection{PULsar Student Exploration online at Parkes (PULSE@Parkes)}

\FloatBarrier
\begin{table}
\caption{PULSE@Parkes Summary}
\begin{center}
\begin{tabular}{@{}lc@{}}
\hline\hline
%Dimension & Classification accuracy \\
%\hline%
Founded & 2007 \\
End Date  & Continuing \\ 
Budget/Yr  & Internally Funded \\ 
Funding Source  & CSIRO Astronomy \\ & and Space Science  \\ 
Cost to participants & Free \\ 
N (Teachers)  & 1-2/School Group \\ 
N (Students)  & 950 Total \\ 
N (Schools)  & 78 Total \\
Location & Australia \\ 
Website & pulseatparkes.atnf.csiro.au \\ 
\hline\hline
Population  & Students \\ 
Inquiry Depth  & Structured/Guided \\ 
Data Rawness  & Newly Observed \\
Selectivity & Not Selective \\ 
Scalable & Not Scalable,\\ & limited by telescope time \\ 
On/Off-campus & External \\
\hline\hline
\end{tabular}
\end{center}
\label{tab1}
\end{table}

PULSE@Parkes (Hollow et al. 2008) was established at CSIRO's Australia Telescope National Facility (ATNF), now CSIRO Astronomy and Space Science (CASS), in late 2007 as a means of engaging high school students in science through radio astronomy with the use of a major national facility. Intrinsic to the program is the opportunity for students to meet and engage with professional astronomers and PhD students. Staffing comprises a Coordinator and Education Lead, Project Scientist and post-doctoral fellow. Other former CSIRO post-doctoral fellows have been active team members, and both PhD students and undergraduate students have taken part in observing sessions or contributed to the development of materials.

Papers have been presented at science teacher and astronomy conferences in Australia and internationally (Hollow et al. 2008). To date, one science paper describing the observing process and data has been published (Hobbs, et al. 2009) whilst another paper based on a specific pulsar is currently being written.

Schools apply for observing sessions that are determined by the schedule for the Parkes radio telescope which is allocated twice per year in six-month blocks. The maximum recommended class size is 24. Prior to the observing session, the project coordinator visits the school to give a one-lesson background talk preparing students for the observations.

On the day, the school group links directly via a Skype link to the astronomer in the tower at Parkes itself. The project catalog has 42 pulsars, selected to provide sufficient coverage at any time of day and year, and that exhibit a variety of pulsar properties. Students split into small groups of 2-4 students; each group has to identify 2-3 pulsars that are up, control the telescope and gather data for the pulsars, and then analyse their data to determine pulsar distance. Observations usually last for two hours and are aided by at least three staff and/or PhD students. 

The project is ongoing and continues to evolve. One of the aims of the project was to inform development of future radio astronomy projects for schools that would utilise the massive data sets expected from new generation radio telescopes, specifically the Australian Square Kilometre Array (ASKAP), operated by CSIRO and the international Square Kilometre Array (SKA).

\FloatBarrier
\subsection{Remote Access Astronomy Project (RAAP)}

\FloatBarrier
\begin{table}
\caption{Remote Access Astronomy Project Summary}
\begin{center}
\begin{tabular}{@{}lc@{}}
\hline\hline
%Dimension & Classification accuracy \\
%\hline%
Founded & 1988, First Light 1992 \\
End Date  & 2003 \\ 
Budget/Yr  & Unknown \\ 
Funding Source  & UCSB grant, NSF EPO \\ 
Cost to participants & Free \\ 
N (Teachers)  & Unknown \\ 
N (Students)  & Unknown \\ 
N (Schools)  & Unknown \\
Location & Global, mainly USA. \\ 
Website & None \\ 
\hline\hline
Population  & Both\\ 
Inquiry Depth  & Structured/Guided \\ 
Data Rawness  & Newly Observed \\
Selectivity & Not Selective \\ 
Scalable & NA \\ 
On/Off-campus & Mixture \\
\hline\hline
\end{tabular}
\end{center}
\label{tab1}
\end{table}

A 14-inch remote access telescope equipped with an impressive array of filters and polarizers, named the Remote Observation Telescope (ROT), was constructed from scratch by undergraduate students from 1988/9 to about 1992 on top of the physics building at UC Santa Barbara. It was used for about 15 years, initially by undergraduates, but later included high schools accessing and requesting CCD image data as well as communicating with each other and the university via a bulletin board system named AstroRAAP. It was decommissioned around the turn of the millenium due to a roof renovation. 

An early grant for computers and a phone line was acquired and one of the project team initially used it in her high school classroom. A high school course for astronomy that was approved for a college-entrance/college-preparatory class was also designed and used for RAAP. Students learned digital image processing in the first semester and in the second semester doing an actual observing project where they would directly go and sit in the lab at UCSB to use the telescope. 

Based around this approach, workshops were given around the country focussed on digital image processing as well as how to use the RAAP telescope itself. Image processing workshops were conducted at professional meetings and conferences around the country. Initially there was no commercially viable software, so the project ended up writing their own. The software used was called 'IMAGINE-32', a professional quality image processing program (Lubin 1992), which could also be used to control a telescope, CCD camera and its filter wheel.

To make the telescope manageable, RAAP accepted lists of object name or RA \& Dec, filters required, and the exposure time. Some amount of human interaction was needed as the dead pointing of the system was not perfect, so a volunteer would come in during the evenings to help out with the process. The telescope was never completely automated, but approximately 50 images a night were taken depending on the exposure time. There was no huge volume of external users, as there were few teachers who had the capacity to dial in and teach the course during the early modem period. Hence, most of the use was by local high schools. 

\FloatBarrier
\subsection{Research Based Science Education (RBSE)/Teacher Leaders in Research Based Science Education (TLRBSE) / Astronomy Research Based Science Education (ARBSE) (NOAO)}

\FloatBarrier
\begin{table}
\caption{Research Based Science Education Summary}
\begin{center}
\begin{tabular}{@{}lc@{}}
\hline\hline
%Dimension & Classification accuracy \\
%\hline%
Founded & 1997 \\
End Date  & 2008 \\ 
Budget/Yr  & 100,000-300,000USD \\ 
Funding Source  & NSF (1st 8 yrs) \& NOAO \\ 
Cost to participants & Free \\ 
N (Teachers)  & 200 Total \\ 
N (Students)  & NA \\ 
N (Schools)  & NA \\
Location & USA \\ 
Website & http://www.noao.edu/\\ &education/arbse/ \\ 
\hline\hline
Population  & Teachers \\ 
Inquiry Depth  & Guided/Coupled \\ 
Data Rawness  & Newly Observed \\
Selectivity & Selective \\ 
Scalable & Scalable \\ 
On/Off-campus & Mixture \\
\hline\hline
\end{tabular}
\end{center}
\label{tab1}
\end{table}

In 1997, a four-year teacher enhancement National Science Foundation grant set up the initial Research Based Science Education (RBSE) project. Its focus was on providing research astronomy experiences to teachers to enable them to bring these into their classrooms to address the issue that teachers may be teaching misconceptions about what science is all about. This continued past the original grant stage and became Teacher Leaders in Research Based Science Education (TLRBSE) until 2005, which added an aspect of teacher leadership development to the training. Then for four years after this period, it became an internal program at the National Optical Astronomy Observatory (NOAO). With budget stress, it was decided that it was not maintainable and the project's last year in this form was 2008. Many of the teachers involved from RBSE were suggested as teachers for the Spitzer Space Telescope Research Program (also described in this paper under NITARP). A form of RBSE for undergrad¬uates is still run by Travis Rector at the University of Alaska.

The original budget was \$200,000 to \$300,000 a year for 20 teachers. The major costs were to run a distance-learning course, to bring the teachers to Kitt Peak for two weeks, and to take the teachers to a National Science Teachers Association (NSTA) meeting, to provide materials for the teachers as well as to pay some consultants. When the grants came to an end and the project became a core project, the budget contracted to about \$100,000 a year. Some members of NOAO staff donated their time as well. The funding rate per teacher was the primary rationale for the NSF in cancelling these styles of grants. In the initial RBSE, around 16 teachers undertook a 180-hour summer workshop involving image processing training, observing with the five major Kitt Peak telescopes with mentoring provided by professional astronomers and educators interspersing the more traditional lecture-type activities. (Rector et al. 2000).

A total of 57 middle and high school teachers undertook the workshop in the first four-year grant with each successive year involving an entirely fresh group of teachers. There was a 15-week distance-learning course that demanded about 10-15 hours a week from teachers. A two week on-campus course was then held in Tucson to help develop their skills, astronomy knowledge and leadership. About a week of this time was spent on Kitt Peak collecting data for their observational projects. Students and teachers could then take their datasets home to the classroom as well as the datasets obtained in previous years. In order to focus the teachers' attention on the professional side of teaching, they would present their work at the annual NSTA meeting approximately two months after the Kitt Peak course.

ImageJ was the software primarily used to examine the images. However a lot of preliminary work (such as image calibration and embedding a WCS) was undertaken by an astronomer. In this sense, teachers never really used completely raw data. Sometimes custom-software was used for particular projects while other projects used Excel spreadsheets to work with the data. 

There were a number of distinct projects. Initially, there were three: a novae search, AGN spectroscopy and a sunspot position project. Further projects developed over time: making recovery observations of asteroids, spectroscopy of variable stars, and a search for high redshift galaxies. The asteroid data were published in the MPC. The Novae search project is currently wrapping up and is in the process of standardising the photometry ready for publication. It has found more novae than any other prior research group. The novae search was double-blind and there were thousands of students involved over the years.

The RBSE also created a journal (http://www.noao.edu/education/arbse/arpd/journal) that ran for the twelve years and contained much of the research undertaken by students. Initially, this was printed but then moved online. As well as the journal, there more than 30 science fair projects. One RBSE participant, Rick Donahue, won the \$20,000 New York Wired Technology Award for his Project SunSHINE (Lockwood, private communication). 

\FloatBarrier
\subsection{Research Experience for Teachers (RET) - NRAO}

\FloatBarrier
\begin{table}
\caption{Research Experience for Teachers Summary}
\begin{center}
\begin{tabular}{@{}lc@{}}
\hline\hline
%Dimension & Classification accuracy \\
%\hline%
Founded & 1999 \\
End Date  & Continuing \\ 
Budget/Yr  & Unknown \\ 
Funding Source  & NSF \\ 
Cost to participants & Free, Paid Stipend \\ 
N (Teachers)  & Unknown \\ 
N (Students)  & NA \\ 
N (Schools)  & NA \\
Location & USA \\ 
Website & http://www.gb.nrao.edu/\\ &epo/ret.shtml\\ 
\hline\hline
Population  & Teachers \\ 
Inquiry Depth  & Guided \\ 
Data Rawness  & Varies \\
Selectivity & Selective \\ 
Scalable & Not Scalable \\ 
On/Off-campus & External \\
\hline\hline
\end{tabular}
\end{center}
\label{tab1}
\end{table}

Research Experience for Teachers, initiated in 1999, is an NRAO eight-week summer research programme located in Socorro NM, Charlottesville VA and Green Bank WV, where high school teachers are matched with astronomy mentors to participate in a research activity over the course of eight weeks. These teachers are provided with a stipend during the period. The teachers are required to develop, document and implement a unit of work based on their experience, assess and evaluate the efficacy of the unit and then disseminate this work through both their teaching peers via the internet and through an AAS poster, to the astronomical community. 

\FloatBarrier
\subsection{Rutgers Astrophysics Institute}

\FloatBarrier
\begin{table}
\caption{Rutgers Astrophysics Institute Summary}
\begin{center}
\begin{tabular}{@{}lc@{}}
\hline\hline
%Dimension & Classification accuracy \\
%\hline%
Founded & 1998 \\
End Date  & 2013 \\ 
Budget/Yr  & 30,000USD \\ 
Funding Source  & Educational Foundation of \\
& America  and then NASA \\ 
Cost to participants & Free \\ 
N (Teachers)  & 6-8/Year, 25 Total \\ 
N (Students)  & 25/Year, 375 Total \\ 
N (Schools)  & 6-8/Year \\
Location & New Jersey area \\ 
Website & http://xray.rutgers.edu/ \\
& asi/asi\_general.html \\ 
\hline\hline
Population  & Both \\ 
Inquiry Depth  & Coupled \\ 
Data Rawness  & Archival \\
Selectivity & Selective \\ 
Scalable & Scalable \\ 
On/Off-campus & External \\
\hline\hline
\end{tabular}
\end{center}
\label{tab1}
\end{table}

The Rutgers Astrophysics Institute (RAI) focuses on the X-ray range of the electromagnetic spectrum rather than the typical optical or radio. It is a free-of-charge year-long research program for high achieving high school students, teachers, and pre-service teachers (Etkina et al. 1999, Etkina et al. 2003) using data obtainable through the High Energy Astrophysics Science Archive Research Centre (HEASARC) (http://heasarc.gsfc.nasa.gov/) which is the data repository for all NASA high-energy missions.

Funding of around \$50,000 per year was originally received from the Educational Foundation of America. After the funding ended, the project tried to get funding from the NSF three times and was rejected for different reasons. Currently the program is running on \$30,000/year funding from NASA. Originally, a small stipend was provided to teachers, but now only snacks are provided. The grant primarily pays for the instructors. Funding ran out in 2013 after 15 years. In the later stages of the project, New Jersey physics teachers have actually run the institute, rather than academic staff.

There were typically six to eight participating schools per year each with two to four students and one teacher. The process started with a four-week, six hours/day, summer school in June where they learn the necessary physics and astronomy background and methodology through an authentic research education methodology built around the ISLE model (Etkina \& Van Heuvelen 2007). After this, an X-ray source of interest, and of previously unknown nature, was picked to be the object of study for the year. The students and teachers met at Rutgers every two months with a culminating year-end conference as well as discussions on an online board. In total, approximately 25-30 teachers in general have been trained via the program and overall 25 students per year (375 in total) have been through the project.

The nature of the sources that the students analyse changes over time as Chandra and other X-ray observatories collect more data, but the fundamentals of the program, learning about physics and astronomy and data within a constructivist inquiry-based framework, remains the same. The typical types of objects observed were quasars, bursters, X-ray binary sources and super-novae remnants. The students gained significant skills in experimental design and modelling in an authentic research situation, improved their raw physics capacity in terms of an advanced physics test and changed their approach to learning about science (Etkina et al. 2003).  

\FloatBarrier
\subsection{Skynet/Project Observe}

\FloatBarrier
\begin{table}
\caption{SkyNet Summary}
\begin{center}
\begin{tabular}{@{}lc@{}}
\hline\hline
%Dimension & Classification accuracy \\
%\hline%
Founded & 2004 \\
End Date  & Continuing \\ 
Budget/Yr  & 50,000USD \\ 
Funding Source  & NSF \\ 
Cost to participants & Free \\ 
N (Teachers)  & 75 Total \\ 
N (Students)  & 5000 Total \\ 
N (Schools)  & 75 Total \\
Location & USA focussed \\ 
Website & http://skynet.unc.edu/ \\ 
\hline\hline
Population  & Both \\ 
Inquiry Depth  & Structured \\ 
Data Rawness  & Newly Observed \\
Selectivity & Non-selective \\ 
Scalable & Scalable \\ 
On/Off-campus & Mixture \\
\hline\hline
\end{tabular}
\end{center}
\label{tab1}
\end{table}

Skynet began in 2004 with funding to build a cluster of six 0.4m telescopes at CTiO for gamma ray burst follow-up. As a lot of time is spent waiting for an actual burst, the non-observing time was to be used for educational purposes, initially in North Carolina. The focus of the educational part of this project is primarily at the undergraduate level, although high school use has also been made. In total, there have been approximately 40,000 users. 

There was a five- year high school project run with these telescopes through the on-campus Morehead Planetarium. Approximately 75 high school teachers have been trained to use the Skynet Interface and who have used it with thousands of North Carolina school students using the 'Project Observe' curriculum materials. These have a heavy image processing focus. Approximately 5000 students used the high school program and about half of the teachers are still active users of the Skynet telescopes. Although no formal evaluation has taken place, students do arrive at the University of North Carolina (UNC) stating that they have used Skynet in high school. Currently, Skynet is in the process of building a middle-school curriculum.

A high school and undergraduate lab course has been developed at UNC. Custom web-based software was created to overcome installation problems on locked down student computers where the number crunching is done on the server rather than on the students' local computers. Astronomy tools were developed to manipulate FITS files as well as a batch aperture-photometry tool based on IRAF. There are also some simple astrometric tools as well as movie-making functions. Observations can be queued online using a web-page which will usually be executed overnight and telescopes can be monitored as they undertake observations. There is also a 20m radio telescope being built into the Skynet system. This telescope has the same backend as professional telescopes but is designed to be used for education as well.  

\FloatBarrier
\subsection{Space to Grow}

\FloatBarrier
\begin{table}
\caption{Space to Grow Summary}
\begin{center}
\begin{tabular}{@{}lc@{}}
\hline\hline
%Dimension & Classification accuracy \\
%\hline%
Founded & 2009\\
End Date  & 2013 \\ 
Budget/Yr  & 800,000AUD \\ 
Funding Source  & Australian Research Council \\ 
Cost to participants & Free \\ 
N (Teachers)  & 80 Total \\ 
N (Students)  & 4000 Total \\ 
N (Schools)  & 37 Total \\
Location & Australia (NSW) \\ 
Website & http://physics.mq.edu.au/\\
& astronomy/space2grow/ \\ 
\hline\hline
Population  & Teachers \\ 
Inquiry Depth  & Varies \\ 
Data Rawness  & Newly and Pre-Observed \\
Selectivity & Non-Selective \\ 
Scalable & Scalable \\ 
On/Off-campus & At School \\
\hline\hline
\end{tabular}
\end{center}
\label{tab1}
\end{table}

The Space to Grow project (Danaia et al. 2013) was a 3 year funded Australian Research Council grant run through Macquarie University and Charles Sturt University with support from LCOGT and three educational jurisdictions in New South Wales, Australia. The focus of the project was to get high school teachers to utilise the two 2 metre Faulkes telescopes in their classrooms to undertake authentic science. Most of the funding was used to provide capacity for teacher release for training as well as an administrator and postdoctoral position. Over the course of the project, nearly one hundred teachers and thousands of students were involved.

Various levels of professional learning were provided in the project with the main model being multiple (3 to 5) workshop days where the teachers undertook the same processes as the students would in class but with deeper reflection on the materials and how to implement them. The actual in-class implementation involved students undertaking various scaffolding activities such as learning about what telescopes are (Project 1) and how optical astronomical images work (Project 2) based on an investigation approach. The main content-area focus was on the lifecycle of stars (Project 3) through performing interpretation of colour magnitude diagrams that the students created in the classroom from their analysis of open clusters. The materials are described in more detail in (Fitzgerald et al. 2014a)

In Projects 1 and 2, students and their teachers were able to submit observation requests for any object that interests them. Most requests were for 'pretty picture' images, but there was a capacity for students to undertake their own authentic research on stellar cluster astronomy building upon the skills they had learned in Project 3. A number of student groups have had their research published (e.g. Fitzgerald et al. 2012, Fitzgerald et al. 2014b) or have publications currently in review with further research currently at various levels of completion. Research on the impact of the project on students' learning and perceptions of science were assessed using a pre/post design, results of which can be seen in (Fitzgerald et al. 2014c). 

\FloatBarrier
\subsection{Summer Science Program (SSP)}

\FloatBarrier
\begin{table}
\caption{Summer Science Program Summary}
\begin{center}
\begin{tabular}{@{}lc@{}}
\hline\hline
%Dimension & Classification accuracy \\
%\hline%
Founded & 1959 \\
End Date  & Continuing \\ 
Budget/Yr  & 500,000USD \\ 
Funding Source  & Fees (45\%), Alumni (45\%),\\& Grants (10\%) \\ 
Cost to participants & 3950USD \\ 
N (Teachers)  & NA \\ 
N (Students)  & 70/year, 1916 Total \\ 
N (Schools)  & NA\\
Location & Global, mainly USA \\ 
Website & http://www.summerscience.org \\ 
\hline\hline
Population  & Students \\ 
Inquiry Depth  & Coupled \\ 
Data Rawness  & Newly Observed \\
Selectivity & Selective \\ 
Scalable & Scalable \\ 
On/Off-campus & External \\
\hline\hline
\end{tabular}
\end{center}
\label{tab1}
\end{table}

The Summer Science Program (SSP) (http://www.summerscience.org/home/index.php) is one of the longest running hands-on astronomy focussed programs in the world. It was  set up by Paul Routly and George Abell in 1959 (Furutani 2001) and currently runs through support from the membership of their alumni association, and is funded by a related endowment. 

The cost to each student is a \$3950 program fee although there is a provision for needs-based financial aid. The typical budget overall is \$500,000 per year, of which the largest cost is for room and board for 86 people for about six weeks. The funding sources are about 45\% from students, about 45\% from individual donations (mostly alumni and their parents), and about 10\% from grants and investment returns. Initially, the students were all from California who came to the Thatcher School in Ojai, California. Now, participants come from all over the world to two campuses, one in Santa Barbara, California, and one in Socorro, New Mexico. There is one permanent staff member running the project.

Entry to the SSP is highly academically selective. Since 2003, there have been 73 students per summer out of approximately 600-700 applicants. Students attend the program for six weeks over the summer holidays. While there are traditional lectures in the program as well as general field trips, the main focus is on a Near-Earth asteroid observing project with a heavy focus on actually computing the orbital determination from scratch. Most of the time students take most of their data from high-end (14$''$) amateur telescopes, although sometimes students have used larger telescopes, both remotely and in person. They are currently now using CCDs, PCs and the Python programming language. In the early years, this was undertaken using photography, mechanical calculators and their human minds. The results of this project are then submitted to the Minor Planet Center (MPC) at CfA. 

\FloatBarrier
\subsection{Telescopes in Education}

\FloatBarrier
\begin{table}
\caption{Telescopes in Education Summary}
\begin{center}
\begin{tabular}{@{}lc@{}}
\hline\hline
%Dimension & Classification accuracy \\
%\hline%
Founded & 1992 \\
End Date  & 2005 (Currently reemerging) \\ 
Budget/Yr  & 120,000USD \\ 
Funding Source  & Unknown \\ 
Cost to participants & Free \\ 
N (Teachers)  & 2-4 Teachers/School \\ 
N (Students)  & Unknown, 100,000 Users/Year \\ 
N (Schools)  & 400+ Total \\
Location & Global \\ 
Website & None currently \\ 
\hline\hline
Population  & Both \\ 
Inquiry Depth  & Not specific \\ 
Data Rawness  & Newly Observed \\
Selectivity & Not selective \\ 
Scalable & Scalable \\ 
On/Off-campus & At School \\
\hline\hline
\end{tabular}
\end{center}
\label{tab1}
\end{table}

The Telescopes in Education (TIE) project began in 1992 by providing remote control access to a refurbished 24-inch telescope, on loan from Caltech, at Mt Wilson, CA. A control system was built and was controlled via a modified version of Software Bisque's The Sky planetarium software and a control system allowing telescopic control via a modem. Later, a 14$''$ telescope at Mt Wilson and an 18$''$ near Golden, Colorado were brought online (Clark 1998).

Most people involved in the project were volunteers. A telescope operator was employed full-time as well as a part-time administrator. As more volunteers accumulated, the three telescopes were operating every dark minute possible; at its peak, about 100,000 students a year were utilising them. TIE, after a respite, is now focussing on getting Australian students into science and technology through universities and government bodies using real telescopes by using graduate students in the high school classroom.

Initially from 1993, schools could book time and use the 24-inch. As well as being able to control the telescope, the schools would also call in on a speakerphone to speak to the operator at the telescope in case something went wrong, or to ask for advice on such things as exposure times as well as rotating the dome. As technology improved and the internet started taking off, the system was modified to become internet controllable.

While teacher recruitment was gained through presenting at the national education conferences, much of the recruitment was primarily word of mouth. There were no firm requirements set for what was to be done in class with the telescope, although guidance was given when requested. Three teacher workshops per year with 15 teachers per class were held at Mt Wilson for those who wanted to know more about how the telescope actually operated. In terms of curriculum material provided, there were 12 projects, each with only two pages of instructions. Teachers would usually have a mentor who could help. Many students  undertook the photometry projects, a few did astrometry, while only three or four schools did spectroscopic work.

Many students won first place in science fairs they entered. One student won a science fair for Los Angeles (CA, USA) county by determining the rotation rate of Vesta. Another student won the Intel Science Award by getting pictures of NEOs simultaneously from two observatories on a very long baseline and computed their distances while comparing these to the Jet Propulsion Laboratory's radar ranging data acquired at the same time. 

\FloatBarrier
\subsection{Towards Other Planetary Systems (TOPS)}

\FloatBarrier
\begin{table}
\caption{Towards Other Planetary Systems  Summary}
\begin{center}
\begin{tabular}{@{}lc@{}}
\hline\hline
%Dimension & Classification accuracy \\
%\hline%
Founded & 1993 (Pilot), 1999 (Main)\\
End Date  & 1995 (Pilot), 2004 (Main) \\ 
Budget/Yr  & Unknown \\ 
Funding Source  & NSF \& Private donors \\ 
Cost to participants & Free \\ 
N (Teachers)  & 20-30/Year \\ 
N (Students)  & 20/Year \\ 
N (Schools)  & 5-10/year \\
Location & USA and Micronesia \\ 
Website & http://www.ifa.hawaii.edu/tops/
 \\ 
\hline\hline
Population  & Both (Mainly Teachers) \\ 
Inquiry Depth  & Guided \\ 
Data Rawness  & Newly Observed \\
Selectivity & Selective \\ 
Scalable & Scalable \\ 
On/Off-campus & External \\
\hline\hline
\end{tabular}
\end{center}
\label{tab1}
\end{table}

Towards Other Planetary Systems (TOPS) was a five-year Teacher Enhancement project that ran from 1999 to 2004. It originated from an earlier pilot project which ran from 1993-1995 (Meech et al. 2000). It primarily provided professional learning to teachers although students were also funded to participate in the program to gain insights into a career in astronomy or science. The unifying theme was the search for 'habitable' worlds, planetary origins and primitive bodies' (Kadooka 2002).

In a three-week intensive workshop, there were active components where the teachers and students used small 8-inch and 10-inch telescopes to undertake nightly observations. These most commonly involved variable star and double star measurements using photomultiplers and CCDs based around the 'Hands-on Astrophysics' curriculum (Mattei et al. 1997). In addition, simple spectroscopes were used. The results of the investigations were provided to the AAVSO. Other activities included measuring the heights of mountains on the Moon using the smaller telescopes. Some students were involved in an observing run on the NASA Infrared Telescope Facility (http://irtfweb.ifa.hawaii.edu/). Other students were involved in the 2002 TOPS program and presented their work at the July 2002 AAS meeting which included the discovery of a new variable star and a remote observing project utilising the 31$''$ Lowell Observatory telescope (Kadooka 2002). This project eventually evolved indirectly into the Hawai'i Student Teacher Astronomy Research Program (HISTAR) project, described above.

\section{Issues Identified in Projects}

During the research for this article, many informal discussions were conducted with various project personnel from most of the investigated projects as well as executing a more standard literature review. In this section, we present a discussion of project issues that commonly arose during the discussions, as well as those apparent in the published literature. 

\subsection{Funding}
It is quite apparent from all of the projects described that sources of funding are a key aspect of long-term project success. Most of the funding for many of the projects came from short-term grants lasting two to four years that had to be renewed, usually with difficulty, as each grant period came to an end. Many projects end up running on shoestring budgets or with volunteer time after the initial grant period expired. Of very recent note, most of the current projects are in turmoil, or have been terminated, due to programmatic uncertainties associated with NASA EPO funding issues.

While a scientific project can be undertaken in a discrete manner and a question addressed in a particular time-period, educational projects need to be funded in the long-term to maximise their impact on students and teachers. The development time for any education project to get to a stable state typically takes a few years before gains are seen in the classroom (e.g. Johnson et al. 2007, Banilower et al. 2007, Suppovitz et al. 2000) after a typical initial negative 'implementation dip' (Hall \& Hord 2010). That time is spent in developing the materials and ironing out any bugs, as well as reaching a substantial fraction of the intended education community. Once this has been achieved, it requires continual resource input to keep the project alive. 

It is still not clear what the best 'business model' is to provide continual and ongoing support and resources to teachers and projects to ensure their continuation beyond finite funding cycles. Those projects that seem to survive are those that tend to be ones that are relatively low budget and run by volunteers, that are funded as a longer-term project through an existing government body, have become self-funded through charging for their services or through philanthropy.

\subsection{Importance of Evaluation}

In terms of educational research and evaluation, overall there is little known about these projects. There are some minor shining lights, but most projects do not present evaluations, and those that do suffer from significant methodological flaws. For instance, some evaluations focussed simple on very short post-hoc qualitative comments from teachers while others reported no estimates of confident intervals for results from quantitative surveys or used anecdotal successful 'case studies' as an indicator of general success. This situation is not unexpected, as Bretones and Neto (2011) state, educational research in astronomy education needs deeper treatments. The usual reason provided for the lack of evaluation is that it was not part of a grant and was not funded as part of the project. 

However, not all forms of evaluation need be as costly as it is commonly claimed. Simple pre/post-test surveys, such as the Astronomy Diagnostic Test (Hufnagel 2000) for content knowledge and the Secondary School Science Questionnaire (Goodrum et al. 2001) for attitudinal change, can be used in quasi-experimental designs (Gribbons \& Herman 1997) to gain understanding of whether the particular educational design achieved their intended results in the short-term. We recommend that evaluation be undertaken in collaboration with an experienced educational evaluator if at all possible and definitely significantly in advance of the commencement of the project itself. A good general introduction to the issues and techniques project evaluation, "The 2010 User-Friendly Handbook for Project Evaluation" (Frechtling et al. 2010) is available from the National Science Foundation. A recommended guide with a much more direct astronomical focus is "Discipline-Based Science Education Research: A Scientist's Guide." by Slater (et al. 2011)

Indeed, evaluation becomes part of the feedback loop that provides evidence for further grant funding applications.  If a project cannot present solid evidence of efficacy, then the likelihood of further grant funding is reduced dramatically. Hence, evaluation is a necessary component that should be built into the structure of any given project.  

Beyond simply acquiring more grant funding, adequate evaluation allows researchers and investigators to learn from the successes and failures of other projects. Without this knowledge, this field is basically relying on anecdotes from researchers to illuminate any future design process. This is fraught with danger as it is well known that such anecodotal or 'expert' evidence suffers from a "pro-innovation bias" (Rogers 2010), where people involved with projects, due to the nature of human socio-psychology, will evaluate and rate their own projects much more highly than they probably truly are. This leads to complacency on the path to improvement of these projects.

Evaluation itself, of course, is dependent upon the goals of a project. Producing a scientific paper out of a project looks very good for the project, especially in the eyes of research astronomers. It is likely, however, that it is only the peak population in any project who can produce such outcomes. Small numbers of such papers are produced by these projects and are likely to be weighted far too heavily as a measure of success with respect to the educational goals of a project. If the paper is taken in the light of the common goals of these projects, viz.,   giving young people an experience of the scientific method and, more generally, a better understanding of science so that they can appraise the nature of science as a human endeavour to help them make valid choices in the future, then these papers alone are not an adequate measure of 'success'.

An estimation of the level of impact on an individual student can be measured at quite reasonable cost in the short-term. However, longitudinal studies (at higher cost of money and time resources) of the educational impact of such projects on students, such as tracking their actual career path, long-term interest in science or general academic performance, are noticeably absent from the evaluation literature.

What is required is adequate educational research into the true impact of such projects on participating students. Even if the teachers are the primary audience for any given program, the overarching goal of the evaluation should be in seeking answers to the question: "How does this project broadly affect the students?".  Some recent work in fields other than astronomy has suggested that having teachers' experience authentic research has a positive effect on their students who were not involved in the research. For example, Silverstein et al. (2009) find that, within their New York state project, students of teachers who have experienced authentic scientific research pass their state science exams at a rate that is 10\% higher than the students of teachers who have not. The effect on students cannot be discovered by mere exit polls about whether teachers enjoyed a project or found that it improved their teaching. Especially as it is fairly clear that teachers themselves suffer from the same pro-innovation bias that project investigators do (Fitzgerald et al. 2014d).

\subsection{Scalability}

In a perfect world, we would like any astronomical educational initiative to reach as many people as possible, but there are significant barriers for any project in attempting to do this. Usually, funding sets project limits. Many projects by necessity are bounded in size by geography and by time. In the geographical vicinity of a project, it is relatively easy to provide on-campus experiences to teachers and students. Once the project expands beyond the local jurisdiction, the necessity for accommodation of participants, or the travel of facilitators/investigators, is at huge cost. Some projects are also limited by access to telescope time or finite access to archives of data.  

Another less apparent requirement for scalability of these projects is staffing. The skill set necessary to provide adequate staffing for these projects is fairly rare. Faculty personnel with the right mixture of astronomical and educational experience can be difficult to find. If the faculty do not exist, then significant training is necessary to develop the experience in interested people. A unique mixture of content knowledge, pedagogical knowledge, charisma, interpersonal skills and organisational skills must be distributed amongst the project personnel. Even when this is available, the amount of communication overhead in dealing with all of the students' and teachers' questions can be quite large.

\subsection{Adequate Focus and Design}

The experiences of personnel in the projects reviewed make it clear that true open inquiry from scratch simply does not work. This is also clear from the projects reviewed, and also from the literature (Alfieri et al. 2011); providing a telescope to teachers and students with no structure seems not to generate preferable outcomes in an educational context. Sufficient thought into how students interact with the data, what they can learn from the meaning of the data and how they will learn to understand the data is necessary.  

Some thought about how students are actually learning to become scientists within the project is also necessary, as students can manipulate the data and even work out the mechanics of certain aspects but may not be gaining a fundamental understanding of how science works in the process. Sufficient scaffolding of this aspect is necessary to get the students authentically involved in the scientific process. 

\subsection{Technology}
A major barrier that prevented teachers from implementing projects was computing requirements. These have varied strongly with what the project was trying to do, but they have also changed dramatically with the passage of time. The earlier projects (1990s) had a near impossible task trying to get teachers to use computers, even in the rare cases where computers actually existed in their classrooms, let alone getting teachers to interact significantly with the data. 

The technology available in schools can be many years behind that available to research scientists and/or even the home. This presents problems to the provision of cutting-edge hardware and software for research purposes. Even for the more recent projects, lack of bandwidth and user friendliness of the required software were significant impediments to their implementation, both of which are coupled with the need for scientific validity of data and data analysis. Even at this point in time, school email fails often and requires participants to use centralized servers like gmail.com. In addition, schools frequently impose limitations on what and how software can be installed on computers that have to be accessed by students. Technology is also a moving target as the software required to undertake astronomical research forms only a small niche market and the software development can move much more slowly than the pace at which technology itself changes. 

\subsection{Authenticity vs Reality}

A naive research scientist, when approaching this style of project, will generally want the project to be as authentic as possible from the ground up. This, however, is not possible in normal educational settings. Teachers have limited time within their school year and within the scope of what they are expected to teach. In addition, few teachers or students have experience in computer programming. Generally, some projects have shown that if teachers and students acquire the observations and undertake the reductions of their raw data, once they got to interpretation and analysis of the data they had both run out of steam and time. At some point, a decision needs to be made about how authentic the research can be compared to the actual plausibility of the project being successfully undertaken by teachers and students within the educational context. The more authentic the science is to be then more support from project personnel and from research astronomers is necessary, the stronger the skill set that needs to be developed, and, a much greater amount of time needs to be devoted to undertake the project. 

\subsection{Teachers as Scientists}

Another problem arises from the fact that teachers are generally not research scientists and have had no authentic experience of the scientific method. Teachers are generally trained to implement 'activity-based' science where discrete class-long activities can be undertaken successfully with a previously known correct answer. They also have little understanding of the complexities of the actual scientific endeavour beyond that which appears in textbooks and, concurrently, can equally be fairly unaware of their lack of knowledge. Hence, teachers can generally undertake a more 'trivial' rather than an 'intended' authentic scientific investigation approach. 
 
A relatively common comment in the informal discussions was that many of the teachers only undertook the measurement aspects which were relatively more cook-book like and 'safe' while deliberately avoiding interpretation or analysis in their approach. This 'trivial' approach is not unexpected though, as there are extreme demands and limitations on what teachers can undertake within their classroom both in terms of the time and the curriculum-related constraints.

\section{CONCLUSION}

In this paper, we have explored 22 major large-scale projects that have attempted to bring authentic astronomy research into the classroom since the enabling dawn of the internet era and the birth of affordable CCD cameras two decades ago. While there have been many successes and obviously many rewarding valuable experiences for the teachers, students, and project personnel involved, the field still has a number of important issues and problems to address before major successes can be seen in the field. 

The major apparent problems are the very distinct lack of evaluation apparent in the field and, where such evaluation exists, it is hard to access, not publicly available and/or it is methodologically problematic. A closely related issue is the funding of such projects. The best business model for these projects is not at all clear, but without adequate proof of efficacy, any attempt to extract the long-term funding necessary for the support of these educational projects to allow measurable success to become apparent is problematic. 

With sufficient concentration on solutions of these two major issues, and addressing the no less important, but secondary, issues of scalability, good design, technology barriers and the realities of the limitations of classrooms, teachers and time, in the future, we may see sustained large-scale classroom research projects emerging with verifiable measurements of their true efficacy and impact on students and teacher practice.

\begin{acknowledgements}
Appreciation goes out to those project personnel who were interviewed or provided important information in the creation of this paper. John Baruch, Richard Beare, Richard Bowden, Bryan Butler, Gilbert Clarke, Rosa Doran, Ryan Dorcey, Eugenia Etkina, Catherine Garland, Katy Garmany, Edward Gomez, Roy Gould, Frederic Hessmann, Suzanne Jacoby, Rick Jenet, Mary-Ann Kadooka, Fraser Lewis, Jeff Lockwood, Philip Lubin, Patrick Miller, Carl Pennypacker, Stephen Pompea, Jordan Raddick, Travis Rector, Dan Reichart, Paul Roche, Rachel Rosen, Phillip Sadler, Jatila van der Veen, Christi Whitworth and Donna Young
\end{acknowledgements}

\begin{appendix}

\end{appendix}

% UNCOMMENT THE LINES BELOW IF YOU WISH TO USE BIBTEX
%\bibliographystyle{apj}
%\bibliography{yourbibfile}

\end{document}